\documentclass[namedreferences,hyperref,optionalrh,solaromanenum]{spr-sola}

\usepackage{graphicx}                    
\usepackage{color}                       
\usepackage{soul}
\usepackage[T1]{fontenc}

\definecolor{darkgreen}{rgb}{0,0.6,0}

\chardef\us=`\_

\usepackage[table,xcdraw]{xcolor}
\begin{document}

\begin{frontmatter}

\title{\titlecase{The Future of Solar modelling: requirements for a new generation of solar models.}}

%

\author[addressref={aff1},email={gbuldgen@uliege.be}]{\inits{G.}\fnm{Gaël}~\snm{Buldgen}\orcid{0000-0001-6357-1992}}
\author[addressref={aff2}]{\inits{G.}\fnm{Gloria.}~\snm{Canocchi}\orcid{0000-0002-4200-9906}}
\author[addressref={aff3}]{\inits{A.}\fnm{Arthur.}~\snm{Le Saux}\orcid{0000-0002-6686-7956}}
\author[addressref={aff4}]{\inits{V.A.}\fnm{Vladimir A.}~\snm{Baturin}\orcid{0009-0000-7642-6786}}
\author[addressref={aff5}]{\inits{R.}\fnm{Regner}~\snm{Trampedach}\orcid{0000-0003-0866-6141}}
\author[addressref={aff4}]{\inits{A.V.}\fnm{Anna V.}~\snm{Oreshina}\orcid{0009-0004-7969-1840}}
\author[addressref={aff4}]{\inits{S.V.}\fnm{Sergey V.}~\snm{Ayukov}\orcid{0009-0001-8859-3570}}
\author[addressref={aff6}]{\inits{J.C.}\fnm{Anil}~\snm{Pradhan}\orcid{XXXXX}}
\author[addressref={aff7,aff8}]{\inits{J.C.}\fnm{Jean-Christophe}~\snm{Pain}\orcid{XXXXX}}
\author[addressref={aff9,aff10}]{\inits{M.}\fnm{Masanobu}~\snm{Kunitomo}\orcid{0000-0002-1932-3358}}
\author[addressref={aff11}]{\inits{T.}\fnm{Thierry}~\snm{Appourchaux}\orcid{XXXX}}
\author[addressref={aff3}]{\inits{R.A.}\fnm{Rafael A.}~\snm{Garc\'\i a}\orcid{0000-0002-8854-3776}}
\author[addressref={aff13}]{\inits{M.}\fnm{Morgan}~\snm{Deal}\orcid{0000-0001-6774-3587}}
\author[addressref={aff1,aff14}]{\inits{N.}\fnm{Nicolas}~\snm{Grevesse}}
\author[addressref={aff1}]{\inits{A.}\fnm{Arlette}~\snm{Noels}\orcid{XXXXX}}
\author[addressref={aff15}]{\inits{J.}\fnm{J{\o}rgen}~\snm{Christensen-Dalsgaard}\orcid{0000-0001-5137-0966}}
\author[addressref={aff10}]{\inits{T.}\fnm{Tristan}~\snm{Guillot}\orcid{0000-0002-7188-8428}}
\author[addressref={aff16}]{\inits{D.}\fnm{Devesh}~\snm{Nandal}\orcid{XXXX}}
\author[addressref={aff12}]{\inits{J.}\fnm{Jérôme}~\snm{Bétrisey}\orcid{0000-0002-6462-3683}}
\author[addressref={aff7}]{\inits{C.}\fnm{Christophe}~\snm{Blancard}\orcid{XXXX}}
\author[addressref={aff17}]{\inits{J.}\fnm{James}~\snm{Colgan}\orcid{XXXX}}
\author[addressref={aff7}]{\inits{P.}\fnm{Philippe}~\snm{Cossé}\orcid{XXXX}}
\author[addressref={aff17}]{\inits{C.J.}\fnm{Christopher J.}~\snm{Fontes}\orcid{XXXX}}
\author[addressref={aff18}]{\inits{L.}\fnm{Ludovic}~\snm{Petitdemange}\orcid{XXXX}}
\author[addressref={aff11}]{\inits{C.}\fnm{Charly}~\snm{Pinçon}\orcid{XXXX}}

\address[id=aff1]{STAR Institute, Universit\'e de Li\`ege, Li\`ege, Belgium.}
\address[id=aff2]{Department of Astronomy, Stockholm University, AlbaNova University Center, SE-106 91 Stockholm, Sweden.}
\address[id=aff3]{Universit\'e Paris Cit\'e, Universit\'e Paris-Saclay, CEA, CNRS, AIM, 91191, Gif-sur-Yvette, France.}
\address[id=aff4]{Sternberg Astronomical Institute, Lomonosov Moscow State University, 119234, Moscow, Russia.}
\address[id=aff5]{Space Science Institute, 4765 Walnut Street, Suite B, Boulder, CO 80301, U.S.A.}
\address[id=aff6]{Ohio State University, Dept. Astronomy, Columbus, OH, 43210, USA.}
\address[id=aff7]{CEA, DAM, DIF, F-91297 Arpajon, France.}
\address[id=aff8]{Université Paris-Saclay, CEA, LMCE, 91680 Bruyères le Châtel, France.}
\address[id=aff9]{Department of Physics, Kurume University, 67 Asahimachi, Kurume, Fukuoka, 830-0011, Japan.}
\address[id=aff10]{Universit\'e C\^ote d'Azur, Observatoire de la C\^ote d'Azur, CNRS,Laboratoire Lagrange, France.}
\address[id=aff11]{Université Paris-Saclay, Institut d'Astrophysique Spatiale, UMR 8617, CNRS, Bâtiment 121, 91405, Orsay Cedex, France.}
\address[id=aff13]{LUPM, Universit\'e de Montpellier, CNRS, Place Eug\`ene Bataillon, 34095 Montpellier, France.}
\address[id=aff14]{Centre Spatial de Liège, Université de Liège, Angleur-Liège, Belgium.}
\address[id=aff15]{Stellar Astrophysics Centre, and Department of Physics and Astronomy, Aarhus University, 8000, Ny Munkegade 120, Aarhus C, Denmark.}
\address[id=aff16]{University of Virginia, Astronomy Building, 530 McCormick Road, P.O. Box 400325, Charlottesville, VA 22904.}
\address[id=aff12]{Department of Physics and Astronomy, Uppsala University, Box 516, SE-751 20 Uppsala, Sweden.}
\address[id=aff17]{Los Alamos National Laboratory, Los Alamos, NM, 87545, USA.}
\address[id=aff18]{LERMA, Observatoire de Paris, PSL Research University, CNRS, Sorbonne Université, Paris, France.}
%
\runningauthor{Buldgen et al. }
\runningtitle{Requirements and wishes for new solar Models}


\begin{abstract}
Helioseismology and solar modelling have enjoyed a golden era thanks to decades-long surveys from ground-based networks such as for example GONG, BiSON, IRIS and the SOHO and SDO space missions which have provided high-quality helioseismic observations that supplemented photometric, gravitational, size and shape, limb-darkening and spectroscopic constraints as well as measurements of neutrino fluxes. However, the success of solar models is also deeply rooted in progress in fundamental physics (equation of state of the solar plasma, high-quality atomic physics computations and opacities, description of convection and the role of macroscopic transport processes of angular momentum and chemicals, such as for example meridional circulation, internal gravity waves, shear-induced turbulence or even convection. In this paper, we briefly outline some key areas of research that deserve particular attention in solar modelling. We discuss the current uncertainties that need to be addressed, how these limit our predictions from solar models and their impact on stellar evolution in general. We outline potential strategies to mitigate them and how multidisciplinary approaches will be needed in the future to tackle them.
\end{abstract}

%


\end{frontmatter}

%
\section{\titlecase{Introduction}}\label{sec:intro} 

Over the last decades, the evolution of solar modelling has been marked by both rapid developments and crises. We refer to the recent review by \citet{JCD2021} for a far more detailed overview than what will be presented here. Helioseismology, with its exceptional data quality, has allowed inferences of more than 95$\%$ in radius of the entire solar structure and 80$\%$ in radius of the solar internal rotation. Simultaneously, continuous operations of neutrino detectors have allowed to characterise in great details the properties of the nuclear fusion operating in the solar core. Yet, two major crises have marked the evolution of solar modelling. First, the so-called solar neutrino problem that lasted for decades, until neutrino oscillations were taken into account in the interpretation of the neutrino observations \citep[for a brief review, see][]{McDonald2016}. Second, the solar abundance/modelling problem that followed the 30$\%$ revision of the solar metallicity in 2005 \citep{Asplund2005}. To this day, this issue is still unsolved, with some groups claiming a higher solar metallicity \citep{Caffau2011,Magg2022}, as found in the 1990s, while recent helioseismic determinations of the solar metallicity \citep{Vorontsov13,BuldgenZ,Buldgen2024,Baturin2024} and detailed spectroscopic analyses based on full 3D models, in Non-Local Thermodynamic Equilibrium (NLTE) for many elements \citep{Asplund2021} confirm the lower abundance of oxygen as well as the lower solar metallicity. While it was hoped that the recently revised photospheric neon abundance \citep{Young2018,Asplund2021}, would alleviate the problem, the overall disagreements between standard solar models, helioseismic and neutrino constraints remain. In this paper, we will describe some "wishes" for potential avenues of progress, both from an observational, methodological, and theoretical point of view. 

\section{\titlecase{Observational Constraints}}\label{sec:Obs}

The three main data sources available for solar modelling are from spectroscopic, helioseismic, and neutrino observations. First, high-resolution spectra have been available for decades, like, for example, the spectra observed from high altitude stations, such as at Kitt Peak \citep{Kurucz1984,Neckel1984, Kurucz2005, Stenflo2015}, or at the Jungfraujoch \citep{Delbouille1973} as well as Göttingen \citep{Reiners2016}. These atlases only concern the centre of the solar disk as well as the Sun seen as a star, in flux. As mentioned hereafter, more center-to-limb spectra would also be extremely useful \citep{Ellwarth2023}. More recently, additional center-to-limb spectra were observed at the Gregory Coudé Telescope at IRSOL \citep{Ramelli2017, Ramelli2019}. Second, helioseismic constraints have also been available for decades, from multiple space-based \citep{Frohlich1995,Scherrer1995,Gabriel1997,Rhodes1997,Hoeksema2018} and ground-based \citep{Fossat1991,Tomczyk1995,Harvey1996,Chaplin1996,Hale2016} instruments. The available datasets go from low order radial modes ($\ell=0$ $n=6$) to high angular degrees of the spherical harmonics describing the oscillation modes, namely up to $\ell\!=\!250$ \citep{Larson2015} or up to $\ell\!=\!1000$ \citep{Reiter2020}. The observed modes are, however, purely acoustic modes, and there has been so far no unambiguous detection of individual solar gravity modes. Finally, neutrino observations have been available well before the first helioseismic inversions were conducted \citep[see][for a review]{Villante2021}, but over the years, the precision of the data increased tremendously and allowed to study the nuclear reactions of the core in great detail. Recently, the Borexino experiment reported the detection of CNO neutrinos \citep{Borexino2020, Appel2022,Basilico2023}. Such a detection is particularly important in the context of the current issues linked with the abundance revision. 

\subsection{\titlecase{Spectroscopy}}\label{sec:Spectro}

A detailed discussion on 3D-NLTE abundance analyses in late-type stars has been provided in \citet{Lind2024}. In this review, they discuss the current modelling problem with respect to the debate between the \citet{Magg2022} and \citet{Asplund2021} abundances. The high-quality observations available for the Sun have made it possible to determine precise photospheric abundances for 62 chemical elements, using atomic and molecular lines. 
For neon, the third most abundant metal after oxygen and carbon, detectable spectral lines are absent in the solar photospheric spectrum. Therefore, we have to use other indicators, such as the solar wind \citep{Heber2009, Burnett2019} as well as the far-ultraviolet spectrum of the quiet solar corona \citep{Landi2015}, and the ultraviolet spectrum of the transition region of the quiet Sun \citep{Young2018}. However, these abundance measurements must be taken with caution, as they are measured far from the relevant photosphere (which we know is well mixed with the whole convection zone). These measurements of neon are \emph{at least} affected by the first ionization potential (FIP) effect \citep{Laming2015} which changes their relation to the actual photospheric abundance, as discussed in detail in \citet{Asplund2021}. An innovative method of helioseismic analysis of the adiabatic exponent \citep{Baturin2024}
focuses on heavy element abundances inside the solar convective zone. This approach provides an estimate of the neon abundance, which agrees with the result of \citet{Young2018} for the ratio of neon over oxygen.


Improvements in solar spectroscopic analyses are in large part coupled to improved atomic data. Among the atomic data, we highlight the need for still more precise and more complete transition probabilities for various elements (e.g., \citealt{Li2023}), as well as the need for more reliable cross sections for the collisions with neutral hydrogen atoms as discussed further below, which are key data needed for performing accurate NLTE analyses \citep{Amarsi2018, Grumer2022}. Computationally efficient, but sufficiently complete atomic structure models also play an important role in the NLTE analyses \citep{Caliskan2024,Ma2024}. We must encourage the groups all over the world that produce these data we need and collaborate with these groups to speed up progress in the field.

A good example of this is found in the current debate around the solar oxygen abundance. Broadly speaking, the solar oxygen abundance can be inferred from three different types of diagnostics, each with their own systematic errors. First, the permitted lines, in particular the 777\,nm triplet, typically show strong departures from LTE and a large sensitivity to the inelastic hydrogen collisions that are not adequately described by current theories \citep{Amarsi2018}. Second, the forbidden lines, in particular the 630\,nm, are insensitive to departures from LTE but are all heavily blended \citep[e.g.,][]{Caffau2013}. Third, the infrared molecular lines of OH molecules have historically been plagued by uncertain molecular data but these data are now much improved \citep{Li2015}. Combined with 3D LTE modelling they may therefore be expected to give a precise and accurate abundance based on many lines in different bands \citep{Amarsi2021}. Departures from LTE remain an uncertainty but there are reasons to expect that NLTE effects are small \citep{Hinkle1975,Ayres1989}. Therefore more weight should be given to the abundance determination from OH lines in the infrared, yielding $8.70\pm0.04$ \citep{Amarsi2018, SchmidtMay2024}.

In the future, 3D-NLTE analyses for CNO-bearing molecules should be performed to verify that departures from LTE are small for the infrared lines studied in e.g. \citet{Amarsi2021}, as a consistency check of their abundances from atomic lines. However, spectrum synthesis in NLTE requires a large set of atomic data, including the oscillator strengths for bound-bound radiative transitions, the cross sections for bound-free radiative transitions and for collisions with electrons, all of which are stored in the so-called "model atom" \citep{Lind2024}. Producing these model atoms for heavy elements and molecules remains a significant challenge due to their complexity and the incomplete and often unreliable data currently available. A recent analysis of 
NLTE in H$^-$, the main contributor to the continuous opacity in the visible, fortunately indicated this to be insignificant \citep{Barklem2024}.

This does not mean that solar abundances of elements with well-constrained atomic data can be considered carved in stone and that further improvements should not be sought. It is true that the inferences strongly depend on the quality of the large number of atomic data required, but the model of the photosphere also plays a critical role. We now have 3D models, such as those from Stagger \citep[published in][]{Galsgaard1995, Magic2013,RodriguezDiaz2023, Stein2024} and CO$^5$BOLD \citep{Wedemeyer2004, Freytag2012, Ludwig2012}, that significantly outperform all the 1D models \citep{Pereira2013}, including 1D models derived from temporal and spatial means of 3D models, known as $\langle \mathrm{3D} \rangle$ models (e.g., \citealt{Magg2022}; \citealt{Zhou2023}). Further comparisons of the center-to-limb variations (CLVs) of spectral lines should also be performed, as done by \citet{Ellwarth2023} for Fe I lines or in \citet{Canocchi2024} for Na I lines, to further probe and validate the 3D-NLTE models currently in use. Additionally, the effects of magnetic fields should be considered \citep{Fabbian2010}, even if others find these effects to be negligible for abundance determinations \citep{Shchukina2015, Shchukina2016} but not on the CLVs \citep{Ludwig2023, Kostogryz2024}. It is clear that progress in these different fields, even small, will allow us to refine the abundances of the elements derived from the analysis of the solar spectrum and reduce the uncertainties of the results.

Overall, the current state of spectroscopic analyses, while well established, still has room for progress:
\begin{itemize}
\item Improvements of the atomic data (transition probabilities, photoionization cross sections, collision cross-section with neutral hydrogen),
\item 3D-NLTE analyses of CNO-bearing molecules and further validation of the 3D-NLTE models from CLVs,
\item Study of the effects of magnetic fields, even if their impact seems to be minimal from studies so far.
\end{itemize}

\subsection{\titlecase{Helioseismology}}\label{sec:Helio}

In the last decades, helioseismic constraints have been the backbone of the validation of solar models. Modern datasets count more than $2000$ individual mode frequencies \citep[e.g.][for the MDI instrument, amongst others]{Korzennik2005,Korzennik2008a,Korzennik2008b,Larson2015}, with some recent publications providing up to $6000$ individual mode frequencies or more \citep{Reiter2020}. In this context, the main game-changer for solar evolutionary modelling would be the unambiguous detection of the elusive solar gravity modes. Various authors have reported detections of either a global signature of the dipolar g-mode pattern \citep[$\Delta P$:][]{Garcia2007} similar to what is done in more evolved stars \citep[e.g.][]{Bedding2011} or individual g modes by \citet{Fossat2017}. So far, there has not been an independent confirmation of the existence of these individual g modes in the Sun  \citep[see e.g.][for a discussion on the most recent claims]{Schunker2018,Scherrer2019,Appourchaux2019,Boning2019}. Detecting them would be of primary importance for mapping the dynamics of the inner solar core, as well as for providing constraints, complementary to neutrino fluxes, on conditions in the solar core.  

The most recent review of paths to g-mode detection can be found in \citet{Appourchaux2013}.  The fact that this publication is more than 10 years old shows the difficulty of progressing in this direction. This review presented two major proposals for g-mode detection: one using the Global Oscillation at Low Frequency New Generation instrument \citep[GOLF-NG:][]{Golf2006}, and one using general relativity for detecting the perturbation in the gravitational potential \citep{Appourchaux2009}, see also \citet{Ni2024} for recent developments.  The long development time for missions aiming at gravitational detection (such as LISA) is another demonstration that a mission such as ASTROD-GW \citep{Selig2013,Ni2013} may only be feasible in the distant future.

The main game-changer in helioseismology would be the detection of the elusive solar gravity modes, as they would provide access to the innermost layers of the Sun. Nevertheless, further comparisons between existing datasets for solar acoustic oscillations still remain important due to the high level of precision involved in solar modelling. 

\subsection{\titlecase{Neutrinos}}\label{sec:Theo}

The current observations of neutrino fluxes offer extremely precise constraints on the solar core. Indeed we benefit from precise measurements of the pp, B, Be, pep fluxes which allow us to know quite precisely the temperature of the solar core, as well as a less precise, but still constraining, measurement of the neutrino flux originating from the CNO cycle. Table \ref{Tab:Neut} summarises the latest reported values of the literature \citep{OrebiGann2021,Appel2022}. Similarly to helioseismic constraints, the high precision of neutrino fluxes raises the question of the impact of systematics in the solar models. This state of affairs advocates for a re-analysis of the seminal paper of \citet{Boothroyd2003}, which determined the required numerical precision and consistency to actually compare standard solar models to the available observations. 

Moreover, with the publication of new fusion reaction rates in \citet{Acharya2024}, it has become desirable to analyse the change induced by these updates on neutrino fluxes. While the precision on the CNO neutrino fluxes is currently low, an important point to mention is that some experiments lead to somewhat conflicting results, for example, the fluxes reported by Borexino do not necessarily match those of meta-analyses like that of \citet{Bergstrom2016} and \citet{OrebiGann2021}. This was discussed in \citet{Salmon2021}, who also discussed the importance of comparing models with various physical ingredients to avoid conclusions solely based on abundances. A promising avenue for more detailed investigations is the use of linear solar models \citep{Villante2010} which may be used to complement seismic models. In light of the existing disagreements between various experiments at a high level of precision, it is important to mention that on top of a refining the CNO neutrino flux detection from Borexino \citep{Borexino2018,Borexino2020}, an independent confirmation would also be immensely valuable for accurately knowing the allowable range of our models. This is of crucial importance given the complementary nature of the CNO neutrino flux to the measured p-p chain fluxes and its sensitivity to the central temperature, central abundances and the formation history of the solar system \citep{Kunitomo2021}.
The screening of nuclear charges by free electrons also play a crucial role in the determination of the central temperature of the Sun, and hence the ratio of p-p to CNO reactions and neutrinos (see D{\"a}ppen, elsewhere in this issue).

Regarding neutrinos, a crucial progress avenue is the independent confirmation of CNO neutrino fluxes measured by the Borexino experiment using an independent setup. Reanalysis of the data by \citet{Appel2022} and \citet{Basilico2023} have provided a higher precision to this value, but independent measurements would be essential to anchor the accuracy of the current values.

%
\begin{table}
\caption{Neutrino fluxes for pp, pep, B, Be and CNO.
}
\label{Tab:Neut}
\begin{tabular}{cc}     
\hline                     
Flux & Value \\
  \hline
$\phi(\rm{pp})$ $\left[ \times 10^{10} \rm{cm}^{-2}\rm{s^{-1}}\right]$ & $5.970^{+0.004}_{-0.003}$  \\
$\phi(\rm{pep})$ $\left[ \times 10^{8}\rm{cm}^{-2}\rm{s^{-1}}\right]$  & $1.448 \pm 0.013$  \\
$\phi(\rm{Be})$ $\left[ \times 10^{9}\rm{cm}^{-2}\rm{s^{-1}}\right]$ & $4.80^{+0.24}_{-0.22}$ \\
$\phi(\rm{B})$ $\left[ \times 10^{6}\rm{cm}^{-2}\rm{s^{-1}}\right]$ & $5.16^{+0.13}_{-0.09}$ \\
$\phi(\rm{CNO})$ $\left[ \times 10^{8}\rm{cm}^{-2}\rm{s^{-1}}\right]$ & $6.6^{+2.0}_{-0.9}$ \\
\hline
\end{tabular}
\end{table}
\section{\titlecase{Macroscopic Physics}}\label{sec:Macro}

Macroscopic phenomena acting in the Sun include convective transport of heat and elements, turbulent and diffusive mixing processes, mass-loss and accretion, differential rotation, meridional circulation, the effects of solar magnetism, etc. Convection has been a major uncertainty in stellar physics for the whole history of the field. The current state of convective modelling in standard solar models is a simple calibration of the mixing length parameter based on global solar properties. This free parameter allows to get a very good approximation of the mean convective flux, but it is also a weakness of the theory as it can absorb any number of inadequacies of the near-surface physics we employ in our solar structure models. In particular, multi-dimensional numerical simulations of solar convection have pointed out how convection is indeed three-dimensional (3D), anisotropic, non-linear and time dependent and disagree in important ways with the computationally far cheaper 1D models\citep[see the review by][]{Kupka2017}. However, further efforts are required on the numerical modelling side as there are unexplained discrepancies with observations. Recent progress has been made on including results of hydrodynamical simulations of the top of the convective envelope, as also used in 3D spectroscopic analyses (see Section \ref{sec:Spectro}), in stellar evolutionary models \citep{trampedach:T-tau,trampedach:alfa-fit,Mosumgaard2018,Jorgensen2018,Spada2018,Manchon2024}.

In deeper convective layers, the so-called Convective Conundrum, a disagreement on the dynamics of large-scale flows between observations and numerical simulations \citep{Proxauf2021, Hotta2023}, leads to much more significant uncertainties. Possible solutions could come from small-scale dynamos \citep{Hotta2022,Vasil2024} or small vortex rings transporting low entropy through the bulk of the convection zone, with minimal mixing into the upflow; this is the entropy rain hypothesis \citep{Brandenburg2016,Anders2019}. Another study has suggested that rotation might also partially explain the discrepancies between numerical simulations and observations \citep{Vasil2021}.

A clear weakness of solar models is their treatment of the radiative-convective interface, with the mixing length theory (MLT) being particularly inadequate in these regions. Historical works have pointed at the need for extra-mixing at the base of the solar convective zone \citep[e.g.][]{Monteiro1994}. Calibrated prescriptions of this extra mixing, as overshooting or penetrative convection, have been explored in recent years \citep[e.g.][]{JCD2011,Zhang2019} as possible solutions to the solar abundance problem. Overshooting that does solve the disagreement in sound speed profile, however, has to extend far deeper than the latest prescriptions from hydrodynamical simulations \citep{Baraffe2022}. These global simulations account for large parts of both the radiative and convective zones and are thus particularly well suited to study the interactions between the two zones, in particular mixing at their interface \citep{Brun2011}. Such simulations are also limited, though, by the expected small scale of the braking by the stable layer, compared to the scale of the full Sun, as well as the large disparity between thermal relaxation time-scales and dynamic time-scales at the bottom of the Sun's convective envelope. Based on such global, interface simulations, different overshooting prescriptions have been derived depending on the measured distribution of overshooting depth. Published works include exponential \citep{Freytag1996,Jones2017}, step \citep{Lecoanet2016}, Gumbel \citep{Pratt2017} and Gaussian \citep{Korre2019} distributions. The work by \citet{Pratt2017} also highlights that it is extreme penetrating plumes (and not the average) that characterise the relevant penetration depth in stars, establishing the long-term mixing profile. However, interpretation of these simulations requires some caution as they need to be run with some modifications of physics in order to deal with the disparity of spatial and temporal scales, mentioned above, that can significantly impact physical processes \citep{Baraffe2021, LeSaux2022}. In addition, global solar simulations have confirmed the theoretical predictions \citep[e.g.][]{Spruit1997} about the crucial role played by the near surface layers in driving convection and convective boundary mixing \citep{Vlaykov2022}. However, extending global simulations up to the photosphere remains one of the main challenges of stellar hydrodynamics modelling, the two main obstacles being the large spatial resolution required at the surface (about  $10^{10}$ cells for a resolution of $(24\,{\rm km})^2$, as has been shown to be adequate for granulation-scale simulations to reproduce a large range of solar observations \citealt{Pereira2013}, with a code that has also passed a range of magneto-hydrodynamic benchmark tests \citealt{Stein2024}), and the very short time-steps (of the order of 0.3\,s, set by the flow speeds at the solar surface) required to resolve the motions. A different treatment of radiative transfer, explicitly accounting for dependencies on wavelength and direction, and a realistic equation of state are also needed.

Another macroscopic mixing unaccounted for in standard solar models is rotation. Current prescriptions for rotation in stellar models follow the shellular approximation. The main issue with rotation-induced turbulence is linked to the fact that classical rotating models are in strong disagreement with the helioseismically inferred rotation profile. In such context, additional processes are required to reproduce the inversion results, such as fossil magnetic fields \citep{Gough1998}, internal gravity waves \citep{Charbonnel2005,Pincon2016} and magnetic instabilities \citep{Spruit2002}. The recent work by \citet{Eggenberger2022} linked the potential effects of angular momentum transport to the observed lithium depletion at the solar surface and increase of the helium abundance in the convective zone. A key discriminant as to whether rotation-induced turbulence is responsible for the lithium depletion is the current observed abundance of beryllium. Indeed, the photospheric Be abundance is sensitive to macroscopic transport at the base of the solar convective zone allowing us to calibrate the efficiency of this mixing, regardless of its origin. A significant depletion of beryllium is expected \cite[e.g.,][]{Eggenberger2022}, however, a very recent measurement of the abundance of Be in the photosphere by \citet{Amarsi2024} leads to a depletion of only 31$\%$. Thus further investigations are required to explain the observed very low depletion of Be and link it with an appropriate physical phenomenon.

In addition to ad-hoc calibrations of macroscopic transport, recent simulations have also demonstrated the occurrence in the radiative layers of the so-called magnetic Tayler instability \citep{Petitdemange2023,Petitdemange2024}, or Tayler-Spruit instability \citep{Spruit2002}, which would potentially allow us to explain the internal rotation of the Sun. Such simulations might allow for new formalisms for the Tayler-Spruit instability to be implemented in stellar evolution codes and then studied using helioseismic constraints. However, the Tayler-Spruit instability seems unable to explain the rotation profile of solar-like stars that have evolved off the main-sequence, i.e., solar-mass subgiants \citep{Cantiello2014,Deheuvels2014,Deheuvels2020}. Therefore, further work is also required on asteroseismic targets to understand the underlying mechanism responsible for angular momentum transport in solar-like stars or whether multiple mechanisms might be at play at different phases. As for the solar case, only the determination of the rotation of the inner solar core will allow us to differentiate between the various suggested candidates in the literature, as so far none of them seem to be able to explain the observations of post-main-sequence stars by asteroseismology, despite revised prescriptions being tested and new inversion techniques being developed to further constrain angular momentum transport processes after the main-sequence using space-based photometric observations (see e.g. \citealt{Fuller2019,Fellay2021, BuldgenMCMC2024}, and \citealt{Aerts2019} for a recent review). 

\section{\titlecase{Microscopic Physics}}\label{sec:Micro}

Microphysical ingredients of solar models include radiative opacities, the equation of state for the solar plasma, the nuclear reaction rates and associated prescriptions for the electronic screening, as well as the formalism for microscopic diffusion. 

Such a long list of physical ingredients cannot be described in exhaustive detail here. However, it is worth mentioning briefly a few desirable avenues of progress for the future, given the importance of these ingredients for both solar and stellar physics. 

Regarding microscopic diffusion, most stellar evolution codes now include it in their computations, and it is considered an intrinsic part of the computations of standard solar models since \citet{JCD1993}. The term ``diffusion'' in this context usually covers three separate processes: gravitational settling, radiative levitation and the actual diffusion of species due to thermal motions. Each of these processes is well understood, but the implications complicate stellar modeling immensely: The metals no longer change in unison, and consequently each element need to be followed separately and equation of state and opacity tables need to be given, and interpolated in, for the order of $5^{(N_{\rm elem}-3)}$ more compositions. This is an insurmountable computational cost and the various available formulations mainly differ by what approximations are made in order to preserve the coupling of metals. However, \citet{Turcotte} showed significant discrepancies in diffusion velocities between various formalisms. In this respect, while the physical process is understood and the equations can be derived accurately from fundamental physics, the treatment in various stellar evolution codes still differs and needs to be tested. Radiative accelerations are certainly the most uncertain ingredient when considering diffusion, due to their direct relation with radiative opacities. However, \citet{Turcotte} showed that their impact at the base of the solar convective zone remained limited. Nevertheless, a revision of this analysis would be desirable in relation to the publication of updated opacity tables. 

As for the equation of state, the most up-to-date current equations of state are the FreeEOS \citep{Irwin}, the SAHA-S  \citep{Gryaznov2004, Gryaznov2006,Baturin2017}, the ChemEOS \citep{Kilcrease2015} and the OPAL \citep{Rogers2002} equations of state. A preview of the new {\it T}-MHD equation of state is given elsewhere in this issue \protect\citep{Trampedach:Sierre2024}. Despite its relatively old age, OPAL still constitutes a key reference for the equation of state of stellar material. In particular, the fact that it was developed from a completely different physical approach than all other available equations of states makes it especially attractive for comparisons. However, the fact that it is only available on some specific grid points and considers a limited number of chemical elements constitutes stringent limitations for this equation of state. OPAL EOS includes (besides H and He) four heavy elements – C, N, O, Ne. The main limitation for this equation of state is the absence of Fe. SAHA-S includes eight heavy elements – C, N, O, Ne, Mg, Si, S, Fe, as well as a large group of molecules while the publicly available data are provided in tabulated form for a prescribed relative mixture. FreeEOS includes 20 heavy elements (C, N, O, Ne, Na, Mg, Al, Si, P, S, Cl, Ar, Ca, Ti, Cr, Mn, Fe, Ni), and is a code for computing the EOS on-the-fly, and hence may be calculated for any desirable abundances. FreeEOS is based on the so-called "chemical picture", but the treatment of bound states and the resulting truncation of partition functions by the plasma environment is handled by a parametrization that can emulate various \emph{ab initio} EOS like the MHD EOS and even the Opal EOS, despite the latter being a physical-picture EOS. The latest equation of state available is the SAHA-S equation of state, which has been specifically designed for solar applications. In recent works, \citet{Vorontsov13}, \citet{VorontsovSolarEnv2014} and \cite{Buldgen2024}, the SAHA-S equation of state seemed to be favoured by helioseismic constraints over FreeEOS and OPAL; nevertheless, further comparisons would be required to understand the physical origins of these differences. In addition, it remains clear that all available EOS lie well outside the uncertainties of helioseismic inversions of $\Gamma_1$ in the convective envelope\citep[see e.g.][]{Vorontsov1991, Elliott1996, Basu1997, Elliott1998, Richard1998, Vorontsov2014} (where the stratification is almost exclusively determined by the EOS), the equation of state is therefore still an open problem \citep[see e.g.][for a review]{JCD1992}.

The equation of state of the plasma for main-sequence stars needs to include a number of complex physical processes. Among these are the perturbation of bound electronic states and ionization limits by the plasma environment, the completeness and classification of bound electronic states, realistic and convergent partition functions from these complete sets of perturbed states, the formation of molecules and their pressure dissociation, the Coulomb interaction of charged particles taking into account quantum effects, partial degeneracy and (weak) relativistic effects for electrons, as well as the trivial thermodynamic contribution of LTE radiation. Coulomb interactions between charged particles and the perturbations by the plasma environment, affecting ionization balances are the most obvious and most uncertain effects in the EOS of the solar interior. Currently, neither of them are described adequately, and interactions between effects of different nature are largely ignored.
Regarding the modelling of Coulomb interaction in the equation of state, they all agree on the first-order term, which is the classical Debye-H\"uckel theory. This term is a negative contribution to the pressure, for example, and with increasing density, will surpass all the other pressure contributions, resulting in an unphysical negative total pressure.
Therefore remedies have been sought to limit the DH term to avoid this. One such remedy is the $\tau$-factor describing the effect of a closest approach of particles with the same electric charge \citep{Graboske1969}.

This was first included in the Mihalas-Hummer-D{\"a}ppen (MHD) EOS \citep{Mihalas1988}, and later in ChemEOS \citep{Hakel2004,Kilcrease2015}, but argued against by \citet{Trampedach2006}. The OPAL equation of state is developed within the framework of the physical picture as an “activity expansion in terms of cluster diagrams” and includes Coulomb terms up to order 5/2 in density \citep{Rogers1973}. The SAHA-S EOS treats the Coulomb contribution in “the Debye-Huckel approximation in a grand canonical ensemble”  \citep{LiKalter1969, Starostin2006} that inherits from the physical picture approach, but in the free energy representation. All of these treatments limit an asymptotic behavior for strongly coupled plasma and exclude negative pressure. The maximum of Coulomb contribution in solar interior is expected in the superficial hydrogen ionization zone at the temperature near $10^{4}$K and reaches a remarkable value up to 10$\%$ of total pressure. Therefore even small deviations in theories can lead to different results. This allows to study this effect using helioseismic techniques. For example, the $\tau$-factor, a quantum diffraction factor (accounting for effects of electrons being wave packets of finite extent), and a factor that includes all higher-order terms of the Coulomb interactions, are analysed elsewhere in this volume \citep{Trampedach:Sierre2024}.

One of the current problems of the modern theory of the equation of state is the description of ionization (in particular hydrogen and helium have been investigated by \citet{Baturin2025}) inside the Sun. Here, several effects are combined: the nonlinear interaction between the ionization of hydrogen and helium (these have the largest non-linear and cross-terms due to their high abundances), the significant Coulomb effect, and the treatment of excited states of atoms and ions in the temperature range inside the convective zone of the Sun. These factors determine the profile of the adiabatic exponent $\Gamma_1=\frac{d \ln P}{d \ln \rho}\vert_{S}$ in the convective zone, and need to be known with an accuracy that matches the accuracy of the observations. The two currently available approaches, based on the partition functions of Planck-Larkin and Starostin-Roerich \citep{Baturin2025}, or the occupation probability formalism based on Stark shifts by micro-field fluctuations of the MHD EOS, may not match how the Sun ionizes the plasma at depth. To improve our description of the solar structure, one avenue might be a smooth transition between the limiting cases of the Planck-Larkin and Starostin-Roerich partition functions. Another avenue is to improve on the electric micro-field distribution functions, providing the Stark shifting of bound states into the continuum, which is the main physical mechanism behind the occupation probabilities introduced by \citet{Hummer:mhd1}. They used a simple approximation to the micro-field distribution that has proved to be inadequate. Indeed, the micro-field (as seen by oxygen ions, for example) results from the contribution of all the ions of all the elements present in the plasma. But this is very difficult to model, as the electric micro-field depends on both the charge of the radiator (the particles that is getting affected by the micro-field), and the plasma environment it finds itself in (plasma-coupling and electron-screening parameter). The latter problem can be addressed by Monte Carlo or molecular dynamics simulations, but is complicated a lot by the very low abundance of metals', requiring a very large number of particles in order to get statistically significant results for relevant mixtures. 

The accuracy of modern EOS seems to be satisfactory for the purposes of modelling main-sequence stars. However, detailed helioseismic analyses require exceptional accuracy in calculating the adiabatic exponent $\Gamma_1$, due to the remarkably low observational uncertainty of only $10^{-4}$. In addition, the analysis requires a robust calculation of partial derivatives of $\Gamma_1$ with respect to chemical composition (i.e. of abundance of elements such as C, N, O, Ne etc.). The current accuracy of helioseismic inversion of $\Gamma_1$ provides very strong constraints, which allows an improved understanding of the physical processes in the plasma.

\citet{Baturin2022} proposed a method to approximate the $\Gamma_1$ profile along the solar adiabat in the conditions where hydrogen and helium are almost fully ionized. Within the framework of this method, $\Gamma_1$ is represented as a sum of $\Gamma_1^{\rm HHe}$ calculated for the pure hydrogen-helium mixture by adding a linear combination of the ionization contributions from individual heavy elements. It is shown that the accuracy of the synthesized $\Gamma_1$-profile is at the level of $5 \times 10^{-6}$ in the lower half of the solar convective zone. The method can thus be used to solve several problems. First, as a verification of the calculation of $\Gamma_1$ for a given population of ionization stages of an element. Second, to model the contributions of heavy elements not included explicitly in the EOS calculation. Third, to explore various assumptions about the ionization mechanisms. Fourth, to obtain an EOS-based abundance estimate of metals, against a seismic determination of the Sun's $\Gamma_1$ profile \citep[see][]{Baturin2024} .

\begin{figure} 
\centerline{\includegraphics[width=0.6\textwidth,clip=]{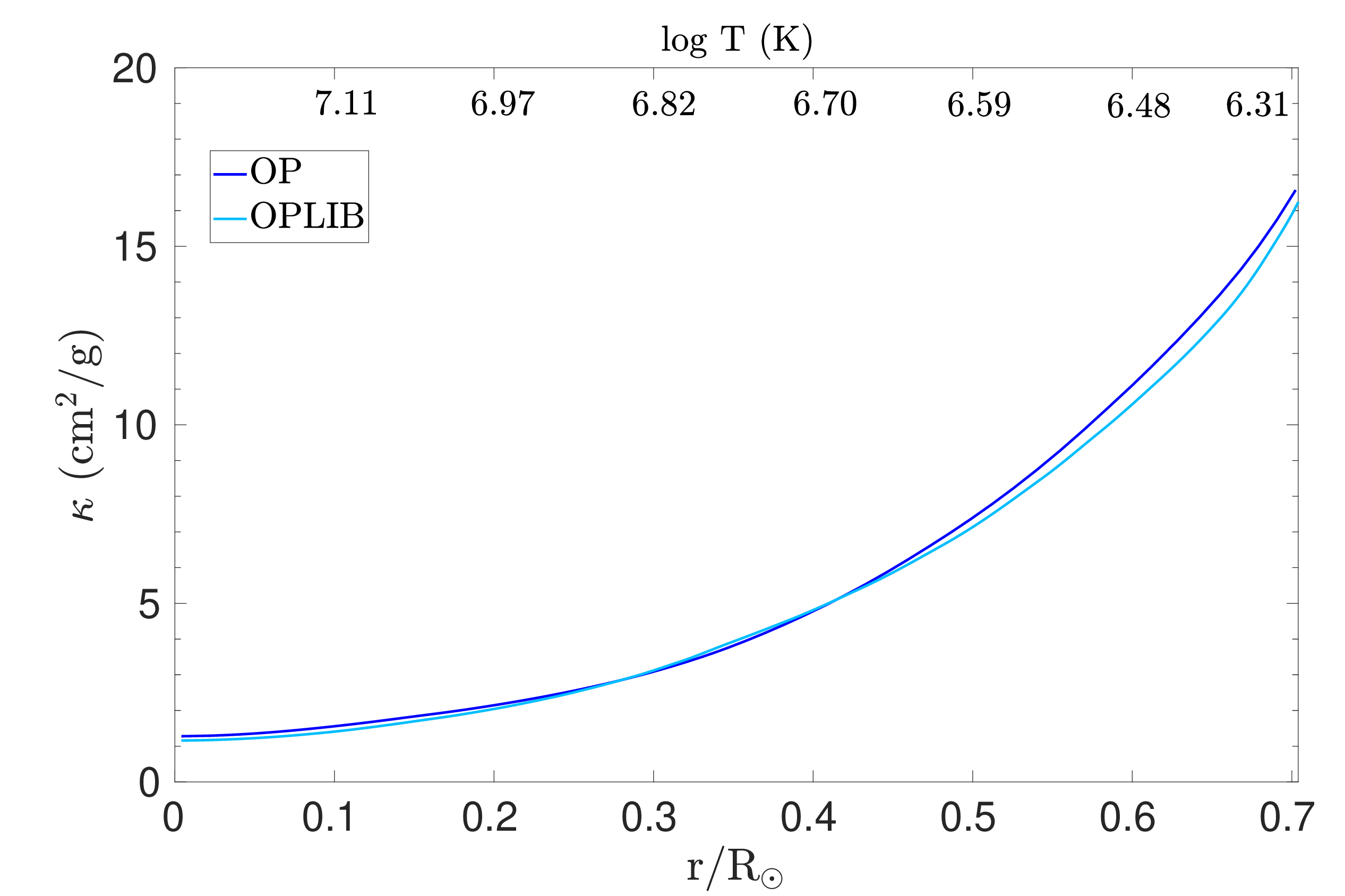}}
 \caption{Comparison between the OP and OPLIB opacities in the solar radiative zone, on the same thermodynamical coordinates ($\rho$, $T$, $X$, $Z$).}\label{fig:OpacComp}
 \end{figure}

The opacities are perhaps the most controversial physical ingredient of solar models, as they have rapidly been identified as the potential source of the solar abundance problem \citep[see, e.g.,][]{Christensen-Dalsgaard2009,Serenelli2009,Trampedach2017b}. Historically, OP \citep{Seaton1994,OP95} and OPAL \citep{OPAL} have been in excellent agreement, but more recent opacity tables such as OPAS \citep{Mondet} and OPLIB \citep{Colgan} have shown more significant deviations. Multiple works have pointed towards the significant differences that have been observed when using the most recent opacity tables  \citep{Mondet,Colgan}. It is worth pointing out, though, that none of them provide a global improvement to the agreement of standard solar models with helioseismic data. Whether one uses OPAS or OPLIB, the sound speed profile is improved at the expense of a lower helium abundance in the convective zone (well outside the helioseismic uncertainty) and, for the OPLIB opacities, a significant disagreement in neutrino fluxes. These effects are due to differences at higher temperatures, at the conditions of the solar core that significantly alter the initial conditions of the solar calibration. See Figure~\ref{fig:OpacComp} for a direct comparison of OP and OPLIB at the same thermodynamical coordinates.

\begin{figure} 
\centerline{\includegraphics[width=0.6\textwidth,clip=]{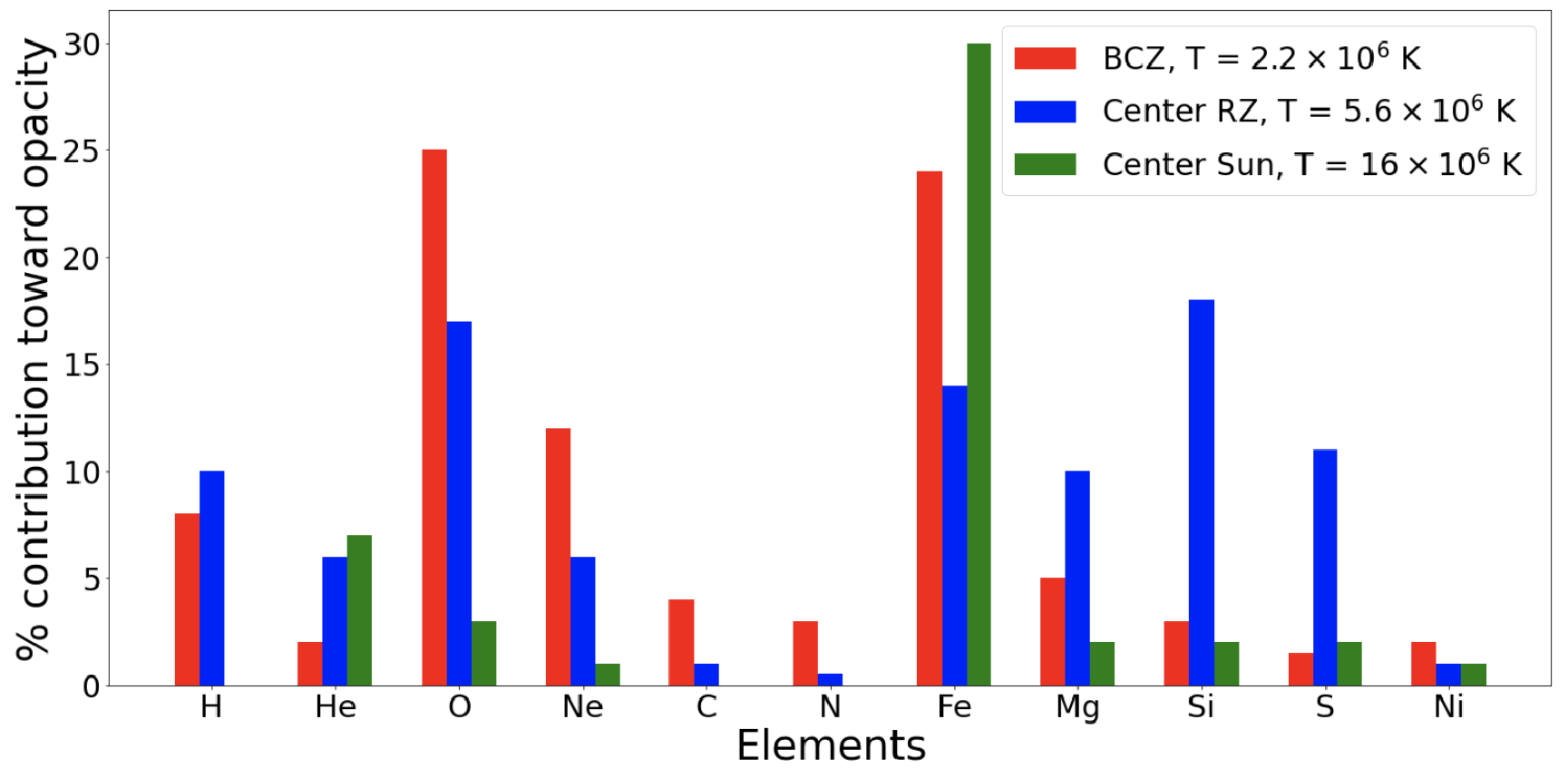}}
 \caption{Contribution of chemical various elements to the Rosseland mean opacity in three characteristic regions of the radiative interior of the Sun.}\label{Fig:OPFe}
 \end{figure}

Numerous works have been produced, recently, following the revision of the solar abundances, as well as the experimental measurement by \citet{Bailey,Nagayama2019,Mayes2025} and the recent experiment by \citet{Hoarty2023}. Theoretical works have also discussed in detail the approximations used in current theoretical opacity computations. Further detailed comparisons between various theoretical opacity codes might give us insight into the origins of the existing discrepancies, while renewed efforts to understand the differences with experimental results might improve the treatment of physical phenomena involved in opacity computations, e.g. Stark effect, number of transitions, auto-ionizing resonances and the underlying equation of state  \citep{Pradhan2024,Pradhan2024R,Nahar2024R,Pradhan2024RII,Zhao2024R}. The issue is particularly acute for elements such as iron which contribute significantly in all regions of the deep solar radiative layers, as illustrated in Figure \ref{Fig:OPFe} for the most significant opacity contributors at three characteristic solar conditions: the base of the convective envelope, the middle of the radiative zone and in the deep core. 

Significant effort needs to be devoted to improving line profiles, in particular the broadening due to ionic Stark effect, for the K-shell lines of intermediate-$\rm{Z}$ elements, especially oxygen. This depends directly on the electric micro-field distributions (also mentioned above for the occupation probabilities in the EOS), and will affect the Rosseland mean opacity by extending line-wings and filling-in opacity-gaps between spectral lines. For heavier elements such as iron, we also need to tackle the modelling of the L-shell Stark effect, but this is a much more ambitious challenge. 

The Z-pinch experiments at Sandia, together with the solar oxygen abundance problem, have focused theoretical efforts on iron L-shell 2p-nd ($n=3$ and 4) transitions. It would be worth investigating the $\Delta n= 0$ (3s-3p and 3p-3d) transitions \citep{DaSilva2002}, which are located under the XUV peak of the Rosseland weighting function at lower temperature ($T \approx 10^{5}$ K), in the convection zone and closer to the surface, in a region also of importance for massive stars \citep{Turck2016}, and are known to be very sensitive to configuration interaction.

It is also essential to make further progress on so-called "density effects", i.e. the impact of the plasma environment on wave functions and hence line energies and intensities. This means taking into account "channel mixing" \citep{Zangwill1980,Dias1997}, which is intimately linked to the configuration interaction and implies that bound-bound (photo-excitation), bound-free (photo-ionization) and free-free (inverse Bremsstrahlung) opacities can no longer be calculated independently before being added together.

Finally, the accuracy vs. completeness balance requires not only to improve the realism of the models, but also to include an increasing number of states in the calculations. Studies of multi-photon absorption should also be pursued, as they are likely to have a significant impact on both opacities and radiative accelerations \citep{MORE201744,PAIN201823,KRUSE201938,Hansen2020,MORE2020100717,KRUSE2021100976}.

Microphysical ingredients are some of the most important aspects of solar models; therefore  progress on these aspects is particularly crucial to the field. We note here a few capital points:
\begin{itemize}
\item Further comparisons between the various implementations of microscopic diffusion in stellar evolution codes are key to determine the reliability of our numerical implementation of this physical process.
\item The various equations of state used in the literature for the solar plasma should be compared in greater detail, not only regarding macroscopic thermodynamical variables, but also the various states and ion fractions predicted for conditions of the solar interior.
\item Further refinements of the physical phenomena included in the current equation of state are also desirable, particularly the description of ionization, the Coulomb interaction and the effects of the plasma environment. Some of these effects might be directly tested using helioseismic data which would drive significant progress on these issues. 
\item Radiative opacities predicted by the main groups in the field should be thoroughly compared in solar conditions, particularly given the mounting evidence for a stark disagreement between theoretical computations and measurements in laboratory and helioseismic techniques. 
\item On the theoretical side, improvements of the description of line profiles, density effects, resonances and testing the inclusion of additional states in the computation might be first steps to explain the observed discrepancies. 
\end{itemize}

\section{\titlecase{Early Evolution}}\label{sec:EarlyEvolution}

Stars form through the gravitational collapse of molecular cloud cores. The second hydrostatic core at the center is called a protostar \citep{Larson1969, Inutsuka2012}. Radiation-hydrodynamic simulations have suggested that the initial mass of a protostar is typically $\sim$0.003\,$M_\odot$ \citep{Vaytet+Haugbolle2017}, and thus accretion from the protosolar disk is essentially important for the early evolution of the Sun. 

Accretion has three effects on the protosolar evolution: compositional, thermal, and rotational effects. The chemical effect of accretion from the protosolar disk has been discussed in a succession of papers as a potential source of enrichment by metals in the solar interior  \citep{Guzik2010,Serenelli2011, Zhang2019}. In a recent series of publications, \citet{Kunitomo2021} and \citet{Kunitomo2022} linked the effects of accretion during the early solar evolution to the formation of the solar system and the so-called pebble wave using the following scenario.

Planet formation theory predicts that dust grains (particularly $\sim$cm-sized ``pebbles'') rapidly accrete onto the proto-Sun due to the frictional force with the disk gas in the early phase leading to a metal-rich accretion, followed by a metal-poor accretion due to dust filtration by a large proto-planet or dust depletion \citep[][and references therein]{Kunitomo2021}. The early high-$Z$ accretion homogeneously increases the metallicity of the fully convective young Sun. Then, the late low-$Z$ accretion decreases the surface metallicity but the radiative core remains metal-rich. Therefore accretion can create a larger compositional gradient in the solar interior. This predicted higher metallicity in the solar core is favorable for reproducing the observed neutrino fluxes (Section\,\ref{sec:Theo}).

Nevertheless, the magnitude of the accretion effect has not been well constrained. 
The history of the accretion metallicity is closely tied to the growth and drift of dust grains, which are highly complicated. Dust growth faces many challenges: bouncing, fragmentation, charge, and radial drift barriers \citep[][and references therein]{Testi2014,Drazkowska2023}. Since the radial drift of grains occurs due to the frictional force with the gas, the gas-disk structure evolution is also crucial (e.g., the efficiency of turbulence affects the vertical settling of grains, and the radial density structure affects the drift velocity). The disk temperature structure (i.e., the location of ice lines) affects the phase transition of elements, which also affects grain growth and drift.
The formation of proto-Jupiter is probably a key factor in the change from the metal-rich to metal-poor accretion through the filtration of grains \citep{Guillot2014}. Recent meteoritic evidence suggests that the Jovian core should have formed rapidly, within 1\,Myr after the Ca-Al-rich inclusions were condensed \citep{Kruijer2020}.
The solar core metallicity can be higher with a longer disk lifetime, because of the shallower surface convective zone of the proto-Sun. Thus, how and when the protosolar disk dispersed are also crucial, but still actively discussed. Recent disk evolution theory suggests that the magnetic disk winds remove gas from the inner disk, while photoevaporation dominates in the late phase \citep{Kunitomo2020}. The birth environment of the Solar System (and thus the radiation field) also matters through external photoevaporation \citep{Guillot2006}.
The magnetism observed in meteorites indicates that the disk lifetime was relatively short \citep[$\simeq$4\,Myr: ][]{Weiss2021}.
All these factors affect the accretion metallicity evolution but remain uncertain.
Constraints on the protosolar disk evolution by theoretical models and observations are highly desirable.

In addition to the chemical effect, accretion also has thermal and rotational effects on the proto-Sun.
A part of the liberated gravitational energy of accreted materials (so-called accretion heat) is carried into the proto-Sun and affects its thermal evolution. Since self-gravitating systems have a negative heat capacity, accretion energy injection delays the increase of the internal temperature of the proto-Sun and thus the development of a radiative core. In this sense, the thermal and chemical evolution of the proto-Sun are tightly linked. This fraction of energy injected into the proto-Sun is generally assumed to be high but remains uncertain \citep{Kunitomo2017}.
The rotational velocity of the proto-Sun sets the initial condition of the rotational history of the Sun, which should affect the macroscopic mixing in the solar interior (Section\,\ref{sec:Macro}). Accretion injects angular momentum, and thus the proto-Sun can rotate rapidly, but observations suggest that young stars rotate much slower than the break-up velocity, probably due to the magnetic star-disk interaction, yet the detailed mechanism is still under debate \citep[e.g.,][]{Bouvier2014, Gallet2019}.

After the accretion phase, the young Sun had a mass larger than its current value. Although the wind mass-loss rate, $\dot{M}$, of the present-day Sun is low ($\dot{M}\simeq2\times10^{-14}\,M_\odot \rm{yr^{-1}}$), both observations \citep[e.g.,][]{Wood2005} and theoretical models \citep[e.g.,][]{Cranmer2011, Suzuki2013, Shoda2023} have suggested that the young Sun produced more intense winds \citep[see also a review by][and references therein]{Vidotto2021}.
Nevertheless, the exact history of $\dot{M}$ is still uncertain. Observationally, \citet{Wood2005} proposed $\dot{M}\propto t^{-2.33}$, while \citet{Vidotto2021} suggested a shallower power law $\dot{M}\propto t^{-0.99}$, where $t$ is time \citep[see also a broken power law in][]{O_Fionnagain2018}.
These empirical laws lead to the solar initial mass $M_{\rm initial}>1.01\,M_\odot$ and $M_{\rm initial}\sim 1.0005\,M_\odot$, respectively.
Mass loss associated with eruptive coronal mass ejections (CMEs) may increase these values \citep{Cranmer2017, Namekata2022}.

Several studies have developed solar models that include mass loss. \citet{Sackmann2003} discussed the implications for the ``faint young Sun paradox'': a more massive young Sun leads to a higher surface temperature of the Earth and Mars because of a higher solar luminosity and smaller semi-major axes. They have also found that even in the models with $M_{\rm initial}=1.07\,M_\odot$, mass loss does not have a significant effect on the sound-speed profile, lithium depletion, and neutrino fluxes \citep[see also][]{Guzik2010}.
We note that \citet{Basinger2024} emphasized the importance of considering mass loss in conjunction with angular momentum evolution. Since the winds magnetically brake the solar rotation, they claimed that $M_{\rm initial}$ should be less than $1.00135\,M_\odot$ to reproduce the solar rotation.

\citet[][see their Figure\,18]{Zhang2019} showed that mass loss can increase the metallicity gradient in the solar interior and thus neutrino fluxes if the composition of solar winds is different from that of the photosphere. Although this is an interesting scenario, the evolution of the composition of solar winds is still unclear. Indeed, the current solar corona has a different composition than the photosphere due to the first-ionization potential (FIP) effect \citep[e.g., depleted in helium;][]{Schmelz2012}, but solar winds do not necessarily have the same composition as the corona, depending on the wind component (fast winds or slow winds).
The histories of the mass-loss rate and the composition of winds are crucially important to evaluate the effects on the solar structure and evolution, and also to understand the planetary surface environment. Further constraints are highly valuable.

As a quick summary, we identify a couple of important hypotheses that need to be further tested regarding the early solar evolution:
\begin{itemize}
\item A more precise estimate of the protosolar disk lifetime is crucial to better quantify the impact of accretion on the solar core composition, with a clear link to the measurements of solar neutrino fluxes.
\item A more precise description of the efficiency with which energy and angular momentum are deposited by accretion is essential as it affects the efficiency of the mixing of chemicals in the early evolution. 
\item A better understanding of the history of mass-loss in the early evolution of the Sun would help test additional effects that might lead to an enhanced composition gradient in solar models and affect both helioseismic and neutrino constraints. 
\end{itemize}

\section{\titlecase{Inference Techniques}}\label{sec:Inferences}

Solar modelling would not have the place it currently holds if it were not for the development of advanced inference techniques of the internal structure and dynamics of the Sun. The validation of standard solar models through their comparisons with helioseismic constraints has been the first step towards a wider use of this modelling framework for solar-like oscillators. In this respect, helioseismology paved the way for asteroseismic modelling of main-sequence solar-like stars. 

\begin{figure} 
 \centerline{\includegraphics[width=1.0\textwidth,clip=]{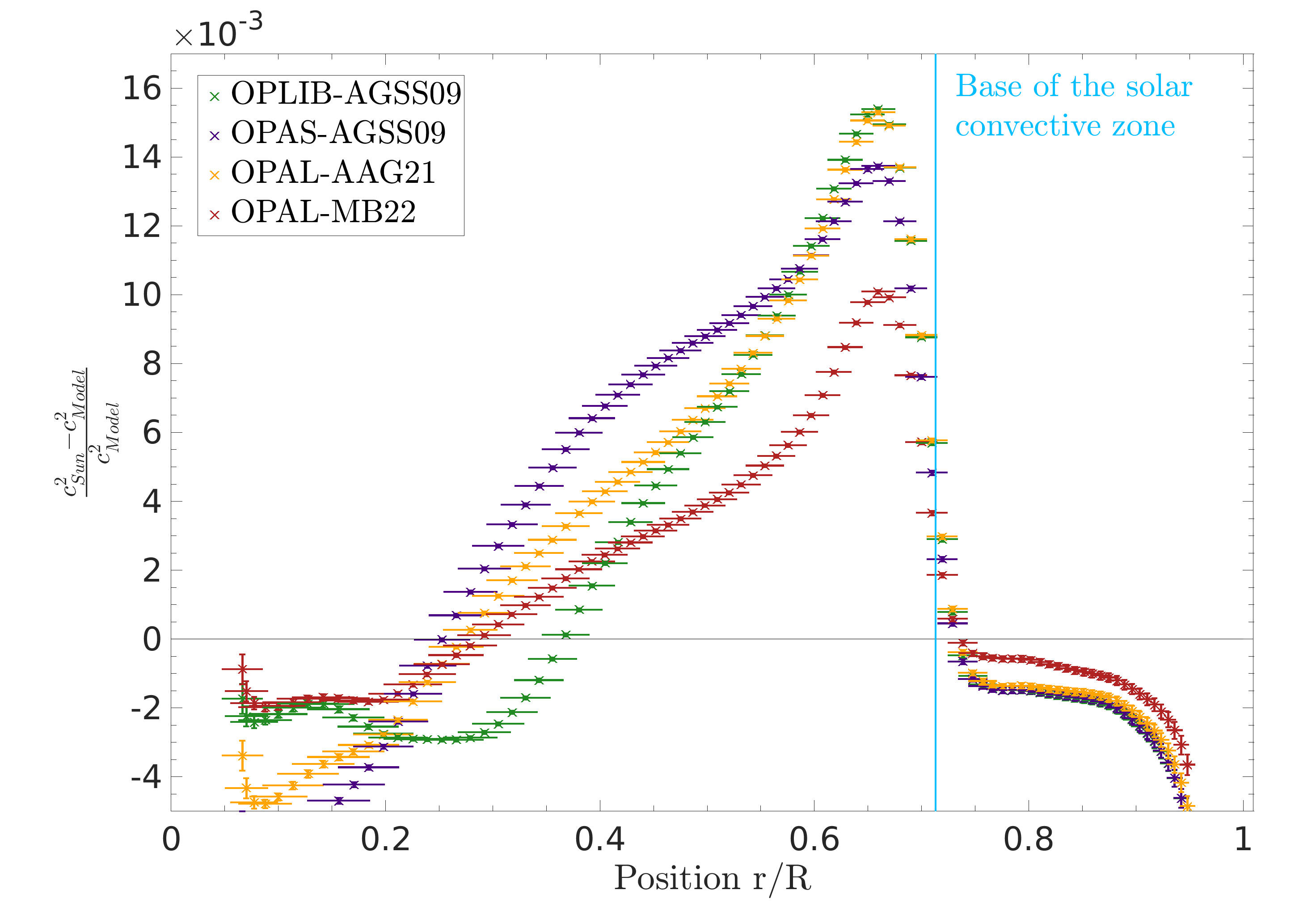}}
\caption{Relative differences in squared adiabatic sound speed for models built with various abundances and opacity tables (namely AAG21, AGSS09 and MB22 for abundances and OPAS, OPLIB and OPAL for opacities).}\label{fig:C2Inversion}
\end{figure}

Nevertheless, standard solar models have also shown their intrinsic limitations that were already noticed in early helioseismic analyses. The most obvious one being the deviation in sound-speed at the base of the convective zone. 
Figure~\ref{fig:C2Inversion} illustrates this issue, whose origin has been linked to multiple causes. Most notably, these include the chemical composition of the solar radiative zone, the opacity of the solar plasma, the effects of the thermalization of convective elements penetrating into these stable layers. This deviation is essentially due to the improper reproduction of the temperature gradient of the Sun by the solar models. Unfortunately, this layer is also the seat of the tachocline, the transition from solid-body rotation to latitudinal differential rotation and \emph{possibly} the location where the solar dynamo originates. Therefore, it is also the region where the 1D, classical, spherically symmetric solar models are second likeliest to departing from the actual Sun--- second only to the top of the convective envelope, located in the middle of the photosphere.

The classical seismic inference techniques used to determine the internal structure of the Sun are linear inversion techniques \citep[see e.g.][]{JCD1990,Kosovichev1999,Buldgen2022} such as the Regularized Least Square \citep[RLS:][]{Phillips1962, Tikhonov1963} or the Optimally Localized Average techniques \citep[OLA:][]{Backus1968,Backus1970,Pijpers1992,Pijpers}. These techniques have, however, intrinsic limitations, with a resolution limit on sharp transitions in both structure and rotation, therefore limiting the conclusions we can draw from actual inversions of sharply varying structural quantities. The inversion results are usually smoother than the actual variations in the structural quantity as a result of the resolution limit of the methods. This implies that processes such as overshooting and its thermalization with the radiatively stratified background medium \citep{Baraffe2022} might require non-linear RLS techniques allowing for solutions with sharper transitions to be resolved, following \citet{Corbard1999}.
Another crucial limitation of inversion techniques is the evaluation of uncertainties. In practice, small differences in global properties (such as the radius of the Sun) might have a significant impact on the conclusions drawn from detailed investigations, as this directly impact the boundary of the integral in the variational expression and might lead to biases inference results. This was recently discussed by \citet{Takata2024} and requires adapting the current inference techniques to their modified variational expressions taking radius variations into account, especially for detailed investigations related to the determination of the solar composition of the convective envelope. Additional stress-tests of the robustness of uncertainty determinations for iterative inversions might be required. This is particularly the case for the treatment of the so-called surface effects, which have amplitudes far exceeding the actual uncertainties in oscillation frequencies \citep{Houdek2017}. Since the inital work by \citet{Rosenthal1999}, some progress has been made in modelling the surface effect based on 3D simulations of convective stellar atmospheres \citet{Ball2016,Trampedach2017a}, although only including the effects due to the convective expansion of the atmosphere, the so-called \emph{structural} component of the surface effect. The interactions between convection and modes, the so-called \emph{modal} component, was tentatively addressed by \citet{Houdek2017}, \citet{Schou2020} and \citet{Trampedach2020}, but is another issue that needs further attention.
The existing literature on magnetic activity is incomplete, and the distinction between activity and surface effects remains ambiguous. Observations have shown that low-degree acoustic frequencies vary with the 11-year solar activity cycle \citep{Woodard1985}, a phenomenon corroborated by multiple studies for low and intermediate modes \citep[e.g.][for a review]{Broomhall2015}. Furthermore, the imprint of the activity cycle of the Sun has been detected in global seismic observables like the large separation \citep{Broomhall2011} and the maximum-power frequency \citep{Howe2020}. Recent findings also indicate that the inferred solar age determined through helioseismology varies significantly with the solar activity cycle \citep{Bétrisey2024}. Given the precision level of helioseismic inversions and the impact of surface effects and magnetic activity on solar frequencies, it is therefore crucial to improve our treatment of the upper convective layers and their interaction with the energetics of the oscillations. In this context, ad-hoc corrections, possibly based on a simple polynomial function of frequency and incorporating additional dependencies on angular degree, maybe inspired by 3D simulations of near-surface layers, are likely to remain useful. Therefore, a systematic analysis of the impact of the biases introduced by such corrections might be required to fully determine how far the detailed inference techniques might be pushed.  

Overall, there is still room for improvement regarding the inference techniques used in helioseismology:
\begin{itemize}
    \item Classical linear inversion techniques should be generalized to non-linear iterative techniques allowing for sharp transitions to be characterised. 
    \item Inclusion of the effects of radius uncertainty and a better estimation of the overall uncertainties of helioseismic inversions. These include comparisons between existing inversion softwares for given datasets and trade-off parameters. 
    \item Further improvements of the descriptions of the so-called surface effects as well as the impact of activity on the inferred results, aiming at providing adequate contingency approaches. 
\end{itemize}

\section{\titlecase{Conclusion}}\label{sec:conc} 

Solar evolutionary models are key calibrators of stellar physics. Over the years, additional physical ingredients, such as microscopic diffusion or revised equations of state, opacities and nuclear reactions rates, have been included in the recipe of what is now the Standard Solar Model. Nevertheless, already in \citet{JCD1996}, \emph{dynamical} processes was mentioned as a potential source for additional improvements of solar models.

Almost three decades later, the physical ingredients of solar models have been significantly updated thanks to numerous theoretical developments. In addition, improvement of spectroscopic and helioseismic data, as well as more precise neutrino fluxes, pave the way for detailed reconstruction techniques. Despite such wealth of observations, however, inferring the solar structure and evolution is still an ill-posed problem, and improving the accuracy of current models will only be achieved through significant progress: 
\begin{itemize}
\item Renewed theoretical opacity computations including plasma broadening effects, background opacities and spectral features that can be directly compared to experimental results. 
\item Renewed molecular dynamics calculations to establish whether screening of nuclear reactions by electrons is static in nature, or dynamic.
\item Independent measurements of CNO neutrino fluxes. Narrowing uncertainties on the neutrino fluxes in general, as neutrinos accumulate.
\item A higher accuracy and precision in the determination of the abundance of key elements, which will play a key role in closing the debate on the solar abundances. 
\item An unambiguous detection of solar gravity modes that can be used to study candidates for angular momentum transport in the solar radiative zone. In this respect, validating the properties of the required transport process using detailed numerical simulations is paramount. 
\item Improved inferences of the sharp transitions in thermal gradients at the base of the convective zone to guide prescriptions for convective overshooting.
\item Improved equation of state calculations with emphasis on including all relevant physical processes, and a goal of completeness in bound states of electrons and in species of particles, i.e., elements, ions and molecules. A few adjustable parameters in the formulation, to be calibrated against helioseismic inversions, might be useful.
\item Forward models of both the structural and modal components of the surface effect are crucial for further progress in helioseismic inversions.
\item  A helioseismic re-determination of the helium abundance, and the neon abundance if possible, based on progress on the EOS and the treatment of the seismic surface effect.
\item  More constraints on the protosolar disk evolution, both from an observational and theoretical point of view are crucial to fully quantify the impact planetary formation may have on the metallicity of the solar core.
\end{itemize}
It is also worth mentioning that the rapid progress of asteroseismology also offers other experimental data points to solar evolutionary models, in the form of solar analogs (i.e. stars of solar mass and composition, but at different ages) and solar twins \citep[see for example the following reviews:][]{Cunha2007, Chaplin2013, Garcia2019, Aerts2019}. The ESA PLATO mission \citep{Rauer2024} will, in this respect, be a game-changer as it will provide hundreds of main-sequence stars for which high quality individual modes will be identified \citep{Goupil2024}. For example, the study of solar twins, as well as the constraints offered by mixed oscillation modes in subgiant and red giant stars, offer complementary insight into potential candidates for angular momentum transport.

%

%
\begin{acks}
We thank the referee for their careful reading of the manuscript and their constructive suggestions. We thank D. Chari for providing Figure \ref{Fig:OPFe}. We thank Dr. A.M.Amarsi for the fruitful discussions associated with the spectroscopic determinations of the solar abundances. GB acknowledges fundings from the Fonds National de la Recherche Scientifique (FNRS) as a postdoctoral researcher. GC acknowledges funds from the Knut and Alice Wallenberg Foundation. RT acknowledges support from NASA grants 80NSSC20K0543 and 80NSSC-22K0829. 
 The study by V.A.B., A.V.O , and S.V.A. is conducted under the state assignment of
Lomonosov Moscow State University.  R.A.G. acknowledges the support from the GOLF and PLATO Centre National D'{\'{E}}tudes Spatiales grants. J.B. acknowledges funding from the SNF Postdoc.Mobility grant no. P$\rm 500PT\_222217$ (Impact of magnetic activity on the characterization of FGKM main-sequence host-stars). M.K. was supported by the JSPS KAKENHI (grant nos. 24K00654 and 24K07099). AP acknowledges partial support from US National Science Foundation (Astronomy). Los Alamos National Laboratory is operated by Triad National Security, LLC, for the National Nuclear Security Administration of US Department of Energy (Contract No. 89233218NCA000001). We acknowledge support by the ISSI team ``Probing the core of the Sun and the stars'' (ID 423) led by Thierry Appourchaux. This research has made use of the Astrophysics Data System, funded by NASA under Cooperative Agreement 80NSSC21M00561.
\end{acks}

%
%
%
%
%
%
%

%
%
\bibliographystyle{spr-mp-sola}
\bibliography{WishlistBiblio}  

\end{document}